\documentclass[%
 reprint,
 amsmath,amssymb,
 aps
]{revtex4-1}
\usepackage{graphicx}% Include figure files
\usepackage{dcolumn}% Align table columns on decimal point
\usepackage{bm}% bold math
\usepackage{color}
\usepackage{amsmath,amssymb,amsfonts}
\input{paperdef}

\graphicspath{{figs/}}

\begin{document}

\preprint{APS/123-QED}

\title{High-precision predictions for the light $\cp$-even 
Higgs Boson Mass of the MSSM}% Force line breaks with \\
%\thanks{A footnote to the article title}%

\author{T.~Hahn$^1$, S.~Heinemeyer$^2$, W.~Hollik$^1$, 
        H.~Rzehak$^3$, G.~Weiglein$^4$}
\affiliation{%
$^1$Max-Planck-Institut f\"ur Physik, 
F\"ohringer Ring 6, D--80805 M\"unchen, Germany\\
$^2$Instituto de F\'isica de Cantabria (CSIC-UC), Santander, Spain\\
$^3$Albert-Ludwigs-Universit\"at Freiburg, 
Physikalisches Institut, D--79104 Freiburg, Germany\\
$^4$DESY, Notkestra\ss e 85, D--22607 Hamburg, Germany}%

\date{\today}% It is always \today, today,
             %  but any date may be explicitly specified

%%%%%%%%%%%%%%%%%%%%%%%%%%%%%%%%%%%%%%%%%%%%%%%%%%%%%%%%%%%%%%%%%%%%%%%%%%%%%
%%%%%%%%%%%%%%%%%%%%%%%%%%%%%%%%%%%%%%%%%%%%%%%%%%%%%%%%%%%%%%%%%%%%%%%%%%%%%

\begin{abstract}

For the interpretation of the signal discovered in the Higgs
searches at the LHC it will be crucial in particular to discriminate between
the minimal Higgs sector realised in the Standard
Model (SM) and its most commonly studied extension, 
the Minimal Supersymmetric SM (MSSM). 
The measured mass value, having already reached the level of a precision
observable with an experimental accuracy of about $500 \mev$, plays
an important role in this context. In the MSSM the mass of the light
$\cp$-even Higgs boson, $\Mh$, can directly be predicted from
the other parameters of the model. The accuracy of this prediction 
should at least match the one of the experimental result. The relatively
high mass value of about 126~GeV has led to many investigations where
the scalar top quarks are in the multi-TeV range. We improve the
prediction for $\Mh$ in the MSSM by combining the existing fixed-order
result, comprising the full one-loop and leading and subleading two-loop
corrections, with a resummation of the leading and subleading
logarithmic contributions from the scalar top sector to all orders.
In this way for the first time a high-precision prediction for the mass
of the light $\cp$-even Higgs boson in the MSSM is possible all the way 
up to the multi-TeV region of the relevant supersymmetric particles.
The results are included in the code \fh.

\end{abstract}

%\pacs{Valid PACS appear here}% PACS, the Physics and Astronomy
                             % Classification Scheme.
%\keywords{Suggested keywords}%Use showkeys class option if keyword
                              %display desired
\maketitle

%\tableofcontents

%%%%%%%%%%%%%%%%%%%%%%%%%%%%%%%%%%%%%%%%%%%%%%%%%%%%%%%%%%%%%%%%%%%%%%%%%%%%%
%%%%%%%%%%%%%%%%%%%%%%%%%%%%%%%%%%%%%%%%%%%%%%%%%%%%%%%%%%%%%%%%%%%%%%%%%%%%%

\section{Introduction}

After the spectacular discovery of a signal in the Higgs-boson searches
at the LHC by ATLAS and CMS~\cite{ATLASdiscovery,CMSdiscovery}, the
exploration of the properties of the observed particle is meanwhile in
full swing. In particular, the observation in the $\ga \ga$ and the
$ZZ^{(*)} \to 4 \ell$ channels made it possible to determine its mass
with already a remarkable precision. Currently, the combined mass
measurement from ATLAS is
$125.5 \pm 0.2 \pm 0.6 \gev$~\cite{ATLAS:2013mma}, and the one
from CMS is $125.7 \pm 0.3 \pm 0.3 \gev$~\cite{CMS:yva}. The other
properties that have been determined so far (with significantly lower
accuracy) are compatible with the minimal realisation of the Higgs
sector within the Standard Model (SM)~\cite{HiggsSMlike}, 
but a large variety of other
interpretations is possible as well, corresponding  to very different
underlying physics. While within the SM the Higgs-boson mass
is just a free parameter,
in theories beyond the SM (BSM) the mass of the particle that is
identified 
with the signal at about $126 \gev$ can often be directly predicted,
providing an important test of the model. The most popular BSM model
is the Minimal Supersymmetric
Standard Model (MSSM)~\cite{mssm}, whose Higgs sector consists of two
scalar doublets accommodating five physical Higgs bosons. In
lowest order these are the light and heavy $\cp$-even $h$
and $H$, the $\cp$-odd $A$, and the charged Higgs bosons $H^\pm$.

The parameters characterising the MSSM Higgs sector at lowest order 
are the gauge couplings, the mass of the $\cp$-odd Higgs boson,
$\MA$, and $\tb \equiv v_2/v_1$,  
the ratio of the two vacuum expectation values. Accordingly, 
all other masses and mixing angles can be predicted in terms of those
parameters, leading to the famous
tree-level upper bound for the mass of the light $\cp$-even Higgs boson, 
$\Mh \le \MZ$, determined by the mass $\MZ$ of the $Z$~boson.
This tree-level upper bound, which arises from the gauge sector,
receives large corrections from the Yukawa sector of the theory, which
can amount up to \order{50\%} (depending on the model
parameters) upon incorporating the full one-loop and the dominant
two-loop contributions~\cite{mhiggsAEC}.
The prediction for the light $\cp$-even Higgs-boson mass in the MSSM is
affected by two kinds of theoretical uncertainties, namely 
parametric uncertainties induced by
the experimental errors of the input parameters, and intrinsic
theoretical uncertainties that are due to unknown higher-order
corrections. Concerning the SM input parameters, the dominant source of
parametric uncertainty is the experimental error on the top-quark mass,
$\mt$. Very roughly, the impact of the experimental error on $\mt$ on
the prediction for $\Mh$ scales like
$\de\Mh^{{\rm para,}\mt}/\de\mt^{\rm exp} \sim 1$~\cite{deltamtILC}. 
As a consequence, high-precision top-physics providing an accuracy on
$\mt$ much below the GeV-level is a crucial ingredient for precision
physics in the Higgs sector~\cite{deltamtILC}. 
Concerning the intrinsic theoretical uncertainties caused by unknown 
higher-order corrections, an overall estimate of 
$\de\Mh^{\rm intr} \sim 3 \gev$ 
has been given in \citeres{mhiggsAEC,PomssmRep} (the more recent
inclusion of the leading \order{\alt\als^2} 3-loop
corrections~\cite{mhiggsFD3l} has slightly reduced this estimated
uncertainty by few \order{100 \mev}), while it was pointed out
that a more detailed estimate needs to take into account the dependence
on the considered parameter region of the model. In particular, the
uncertainty of this fixed-order prediction is expected to be much larger 
for scalar top masses in the multi-TeV range. This region of the
parameter space has received considerable attention recently, partly
because of the relatively high value of $\Mh \approx 126\gev$, which
generically requires either large stop masses or large mixing in the
scalar top sector, and partly because of the limits from searches for 
supersymmetric (SUSY) particles at the LHC.
While within the general MSSM the lighter scalar 
superpartner of the top quark is allowed to be relatively light (down to
values even as low as $\mt$), both with respect to the direct searches 
and with respect to the prediction for $\Mh$ (see
e.g.\ \citere{Mh125}), the situation is different in more constrained
models. For instance, global fits in the Constrained MSSM (CMSSM) prefer
scalar top masses in the multi-TeV range~\cite{mc8,fittino}. 

Here we present a significantly improved prediction for the
mass of the light $\cp$-even Higgs boson in the MSSM, which is expected
to have an important impact on the phenomenology in the
region of large squark masses and on its confrontation 
with the experimental results.

%%%%%%%%%%%%%%%%%%%%%%%%%%%%%%%%%%%%%%%%%%%%%%%%%%%%%%%%%%%%%%%%%%%%%%%%%%%%%
%%%%%%%%%%%%%%%%%%%%%%%%%%%%%%%%%%%%%%%%%%%%%%%%%%%%%%%%%%%%%%%%%%%%%%%%%%%%%

\section{Improved prediction for $\bm{\Mh}$}

In the MSSM with real parameters (we restrict to this case for
simplicity; for the treatment of complex parameters see
 \citeres{mhcMSSMlong,mhcMSSM2L} and references therein),
using the Feynman diagrammatic (FD) approach,
 the higher-order corrected  
$\cp$-even Higgs boson masses are derived by finding the
poles of the $(h,H)$-propagator 
matrix. The inverse of this matrix is given by $-i \times$
\begin{align}
\ML  p^2 -  \mhtree^2 + \hSi_{hh}(p^2)  &  \hSi_{hH}(p^2) \\
     \hSi_{hH}(p^2) & p^2 -  \mHtree^2 + \hSi_{HH}(p^2) \MR,
\label{higgsmassmatrixnondiag}
\end{align}
where $m_{h,H, {\rm tr}}$ denote the 
tree-level masses, 
and $\hSi_{hh,HH,hH}(p^2)$ are the renormalized Higgs boson
self-energies evaluated at the squared external momentum~$p^2$
(for the computation of the leading contributions to those
self-energies it is convenient to use the basis of the fields $\phi_1$,
$\phi_2$, which are related to $h$, $H$ via the (tree-level)
mixing angle $\al$: 
$h = - \sin\al \, \phi_1 + \cos\al \, \phi_2$, 
$H = \cos\al \, \phi_1 + \sin\al \, \phi_2$).
The status of higher-order corrections to these self-energies
is quite advanced. The complete one-loop
result within the MSSM is known~\cite{ERZ,mhiggsf1l}.
The by far dominant one-loop contribution is the \order{\alt} term due
to top and stop loops ($\alt \equiv h_t^2 / (4 \pi)$, $h_t$ being the
top-quark Yukawa coupling). The computation of the two-loop corrections
has meanwhile reached a stage where all the presumably (sub)dominant
contributions are available, see \citere{mhiggsAEC} and references therein.
The public code \fh~\cite{feynhiggs,mhiggslong,mhiggsAEC,mhcMSSMlong} 
includes all of the above corrections, where the on-shell (OS) scheme for the
renormalization of the scalar quark sector has been used (another
public code, based on the Renormalization Group (RG) improved
Effective Potential, is {\tt CPsuperH}~\cite{cpsh}).
A full 2-loop effective potential calculation %
(supplemented by the momentum dependence for the leading
pieces and the leading 3-loop corrections) has been
published~\cite{mhiggsEP5}. However, no computer code 
is publicly available. Most recently another leading 3-loop
calculation at \order{\alt\als^2} became available (based on a
\DRbar\ or a ``hybrid'' renormalisation scheme for the scalar top
sector), where the numerical
evaluation depends on the various SUSY mass hierarchies~\cite{mhiggsFD3l}, 
resulting in the code {\tt H3m} (which
adds the 3-loop corrections to the \fh\ result).

\smallskip
We report here on an improved prediction for $\Mh$ where we combine the
fixed-order result obtained in the OS scheme with an all-order
resummation of the leading and subleading contributions from the scalar
top sector. We have obtained the latter from an analysis of the 
RG Equations (RGEs) at the two-loop level~\cite{SM2LRGE}.
Assuming a common mass scale $\MS = \sqrt{\mste\,\mstz}$ 
($m_{{\tilde t}_{1,2}}$ denote the two scalar top masses, and $\MS \gg \MZ$)
for all relevant SUSY mass parameters,
the quartic Higgs coupling $\la$ can be evolved
via SM RGEs from \MS\ to the scale $Q$ 
(we choose $Q = \mt$ in the following)
where $\Mh^2$ is to be evaluated
(see, for instance, \citere{mhiggsBSE} and references therein),
\begin{align}
\Mh^2 = 2 \la(\mt) v^2~.
\label{Mh2RGE}
\end{align}
Here $v \sim 174 \gev$ denotes the vacuum expectation value of the SM. 
Three coupled RGEs are relevant for this evolution, the ones for
$\la$, $h_t$ %(the top Yukawa coupling, $\alt = h_t^2/(4\,\pi)$) 
and $g_s$ (the strong coupling constant, $\als = g_s^2/(4\,\pi)$). 
Since SM RGEs are used, the relevant parameters are given in the
\MSbar\ scheme.
We incorporate the one-loop threshold corrections to
$\la(\MS)$ as given in \citere{mhiggsBSE}, with
$\xt = \Xt/\MS$, $h_t=h_t(M_S)$,
\begin{align}
\la(\MS) = (3\,h_t^4)/(8\,\pi^2) \xt^2 \KKL 1 - 1/12\, \xt^2 \KKR~, 
\label{threshold}
\end{align}
where as mentioned above $\Xt$ is an \MSbar\ parameter. 
\refeq{threshold} ensures that \refeq{Mh2RGE} consists of the ``pure loop
  correction'' that will be denoted $(\De\Mh^2)^{\rm RGE}$ below.
Using RGEs at two-loop order~\cite{SM2LRGE}, 
including fermionic contributions from the top sector only, 
leads to a prediction for the corrections to $\Mh^2$ including 
leading and subleading logarithmic contributions $L^n$ and $L^{(n-1)}$
at $n$-loop order ($L \equiv  \ln(\MS/\mt)$),
originating from the top/stop sector of the MSSM.
We have obtained both analytic solutions of the RGEs up to the $7$-loop
level as well as a numerical solution incorporating the leading and
subleading logarithmic contributions up to all orders.
In a similar way in \citere{mhiggsRGE3l} the leading logarithms at
3- and 4-loop order have been evaluated analytically.

\smallskip
Concerning the combination of the higher-order logarithmic contributions
obtained from solving the RGEs with the 
fixed-order FD result implemented in \fh\ comprising corrections up to the
two-loop level in the OS scheme, we have used the parametrisation of the
FD result in terms of the running top-quark mass at the scale $\mt$, 
$\overline{\mt} =
\mt^{\rm pole}/(1 + 4/(3 \pi) \als(\mt^{\rm pole}) 
-1/(2 \pi) \alt(\mt^{\rm pole}))$, 
where $\mt^{\rm pole}$ denotes the top-quark pole mass. 
Avoiding double counting of the logarithmic contributions up to the
two-loop level and consistently taking into account the different
schemes employed in the FD and the RGE approach, the correction
$\De\Mh^2$ takes the form
\begin{align}
\De\Mh^2 &= (\De\Mh^2)^{\rm RGE}(\Xt^{\MSbar}) 
          - (\De\Mh^2)^{\rm FD, LL1,LL2}(\Xt^{\OS}) \non \\
\Mh^2 &= (\Mh^2)^{\rm FD} + \De\Mh^2~.
\label{eq:combcorr}
\end{align} 
Here $(\Mh^2)^{\rm FD}$ denotes the fixed-order FD result,
$(\De\Mh^2)^{\rm FD, LL1, LL2}$ are the logarithmic contributions up to
the two-loop level obtained with the FD approach in the OS scheme, while 
$(\De\Mh^2)^{\rm RGE}$ are the leading and sub-leading logarithmic
contributions (either up to a certain loop order or summed to all
orders) obtained in the RGE approach, as evaluated via \refeq{Mh2RGE}. 
In all terms of
\refeq{eq:combcorr} the top-quark mass is parametrised in terms of
$\overline{\mt}$;
the relation between $\Xt^{\MSbar}$ and $\Xt^{\OS}$ is given by
\begin{align}
\Xt^{\MSbar} = \Xt^{\OS} \left[1 + 
2 L
\left(\als/\pi - (3 \alt)/(16\pi)\right)\right]
\end{align}
up to non-logarithmic terms, and there are no logarithmic contributions 
in the relation between $\MS^{\MSbar}$ and $\MS^{\OS}$. 

Since the higher-order corrections beyond 2-loop order
have been derived under the assumption
$\MA \gg \MZ$, to a good approximation these corrections
can be incorporated 
as a shift in the prediction for the $\phi_2\phi_2$ self-energy 
(where $\De\Mh^2$ 
enters with a coefficient $1/\SQb$). In this way
the new higher-order contributions enter not only the prediction for
$\Mh$, but also all other Higgs sector observables that are evaluated in 
\fh. The latest version of the code, \fh\,{\tt 2.10.0}, which is
available at {\tt feynhiggs.de}, contains those improved
predictions as well as a refined estimate of the theoretical
uncertainties from unknown higher-order corrections. Taking into account
the leading and subleading logarithmic contributions in higher orders
reduces the uncertainty of the remaining unknown higher-order
corrections. Accordingly, the estimate of the uncertainties arising from
corrections beyond two-loop order in the top/stop sector is adjusted
such that the impact of replacing the running top-quark mass by the pole
mass (see \citere{mhiggsAEC}) is evaluated only for the non-logarithmic
corrections rather than for the full two-loop contributions implemented
in \fh.
Further refinements of the RGE resummed result are possible, in
particular extending the result to the case of a large
splitting between the left- and right-handed soft
SUSY-breaking terms in the scalar top sector~\cite{mhiggsRGEsplitLR} and
to the region of small values of $\MA$ (close to $\MZ$) as well as including
the corresponding contributions from the (s)bottom sector. We leave
those refinements for future work.

%%%%%%%%%%%%%%%%%%%%%%%%%%%%%%%%%%%%%%%%%%%%%%%%%%%%%%%%%%%%%%%%%%%%%%%%%%%%%
%%%%%%%%%%%%%%%%%%%%%%%%%%%%%%%%%%%%%%%%%%%%%%%%%%%%%%%%%%%%%%%%%%%%%%%%%%%%%

%\mbox{}\vspace{0.3em}
\vspace*{-1ex}

\section{Numerical analysis}

%%%%%%%%%%%%%%%%%%%%%%%%% F I G U R E %%%%%%%%%%%%%%%%%%%%%%%%%%%%%%%%%%%%%%%%%
\begin{figure}[htb!]
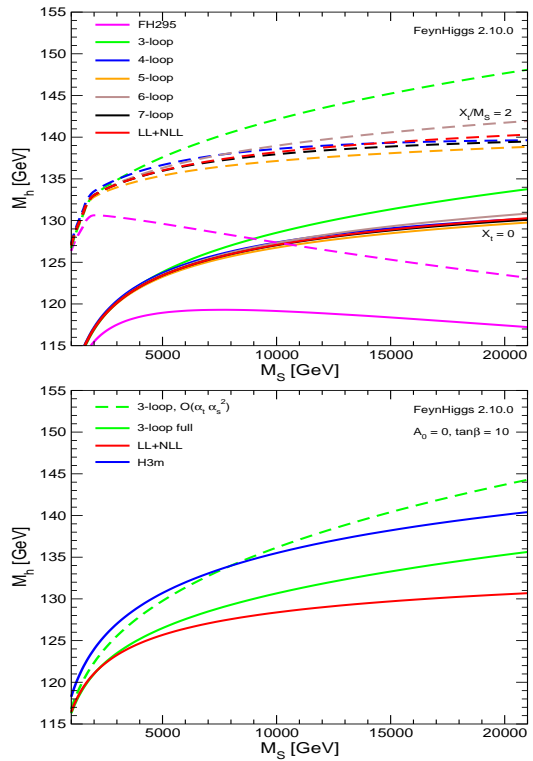

\includegraphics[width=0.385\textwidth,height=5cm]{Mh-logresum-11-Mstop}\\
\includegraphics[width=0.385\textwidth,height=5cm]{Mh-logresum-15-Mstop}
\caption{
Upper plot: $\Mh$ as a function of $\MS$ for $\Xt = 0$ (solid) and 
$\Xt/\MS = 2$ (dashed). The full result (``LL+NLL'') is compared with
results containing the logarithmic contributions up to the 3-loop,
\ldots 7-loop level and with
the fixed-order FD result (``FH295'').
Lower plot: comparison of \fh\ (red) with {\tt H3m} (blue). 
In green we show the \fh\ 3-loop result at \order{\alt\als^2} (full) as
dashed (solid) line.
}
\label{fig:plots}
\end{figure}
%%%%%%%%%%%%%%%%%%%%%%%%% F I G U R E %%%%%%%%%%%%%%%%%%%%%%%%%%%%%%%%%%%%%%%%%

In this section we briefly analyze the phenomenological implications of
the improved $\Mh$ prediction for large stop mass scales, as
evaluated
with \fh\,{\tt 2.10.0}.
The upper plot of \reffi{fig:plots} shows $\Mh$ as a function of 
$\MS$ for $\Xt = 0$ and $\Xt/\MS = 2$ (which corresponds to the minimum
and the maximum value of $\Mh$ as a function of $\Xt/\MS$, respectively;
here and in the following $\Xt$ denotes $\Xt^{\OS}$).
The other parameters are $\MA = M_2 = \mu = 1000 \gev$,
$\mgl = 1600 \gev$ ($M_2$ is the SU(2) gaugino mass term, $\mu$ the
Higgsino mass parameter and $\mgl$ the gluino mass) and $\tb = 10$.
The plot shows for the two values of $\Xt/\MS$ the
fixed-order FD result containing corrections up to the two-loop level
(labelled as ``FH295'', which refers to the previous version of the code
\fh) as well as the latter result supplemented with the analytic
solution of the RGEs up to the 3-loop, \ldots 7-loop level 
(labelled as ``3-loop'' \ldots ``7-loop''). The curve labelled as
``LL+NLL'' represents our full result where the FD contribution is
supplemented by the leading and next-to-leading logarithms summed to all
orders.
One can see that the impact of the higher-order logarithmic
contributions is relatively small for $\MS = \order{1 \tev}$, while
large differences between the fixed-order result and the improved
results occur for large values of $\MS$.
The 3-loop logarithmic
contribution is found to have the largest impact in this context, but 
for $\MS \gsim 2500 (6000) \gev$ for
$\Xt/\MS = 2 (0)$ also contributions beyond 3-loop are important.
A convergence of the higher-order logarithmic contributions towards the
full resummed result is visible.
At $\MS = 20 \tev$ the 
difference between the 7-loop result and the full resummed result is
around $900 (200) \mev$ for $\Xt/\MS = 2 (0)$. The corresponding
deviations stay below $100 \mev$ for $\MS \lsim 10 \tev$.
The plot furthermore shows that for $\MS \approx 10 \tev$ (and the value
of $\tb = 10$ chosen here) a predicted value of $\Mh$ of about $126
\gev$ is obtained even for the case of vanishing mixing in the scalar
top sector ($\Xt = 0$). Since the predicted value of $\Mh$ grows further
with increasing $\MS$ it becomes apparent that the measured 
mass of the observed signal, when interpreted as $\Mh$, can
be used (within the current experimental
and theoretical uncertaintes) to derive an {\em upper bound\/} on the
mass scale $\MS$ in the scalar top sector,
see also \citere{StopUpperLimit}.

In the lower plot of \reffi{fig:plots} we compare our result with the
one based on the code {\tt H3m}~\cite{mhiggsFD3l}
using a CMSSM scenario
with $m_0 = m_{1/2} = 200 \gev \ldots 15000 \gev$, 
$A_0 = 0$, $\tb = 10$ and $\mu > 0$. The spectra were generated with 
{\tt SoftSusy\,3.3.10}~\cite{softsusy}.
The {\tt H3m} result (blue line) is based on the \fh\ result
up to the two-loop order and incorporates the \order{\alt\als^2}
corrections containing also non-logarithmic contributions. Besides our 
result where \fh\ is supplemented by the leading and subleading logarithmic
corrections to all orders (red line) we also show the expansion of our
result up to the 3-loop level (green solid line), containing at this
level the $L^3$ and $L^2$ terms, and the result restricting the
contributions at the 3-loop level to the ones of \order{\alt\als^2}
(green dashed). We find that the latter result agrees rather well with
{\tt H3m}, with maximal deviations of \order{1 \gev} for $\MS \lsim 10
\tev$. The observed deviations can be attributed to the $L^1$ and
$L^0$ terms contained in {\tt H3m}, to the various SUSY mass
hierarchies taken into account in {\tt H3m}, and to the different
renormalization schemes employed. However, one can see that the 3-loop
contributions beyond the \order{\alt\als^2} terms, i.e.\ corrections of 
\order{\alt^2\als, \alt^3} that are not contained in {\tt H3m}, have a sizable
effect giving rise to a (downward) shift in $\Mh$ by $\sim 5 \gev$
for $\MS = 10 \tev$. The corrections beyond the 3-loop order yield an
additional 
shift of about $2 \gev$ for $\MS = 10 \tev$, in
accordance with our analysis above. Larger changes are found for $\MS >
10 \tev$.

%%%%%%%%%%%%%%%%%%%%%%%%%%%%%%%%%%%%%%%%%%%%%%%%%%%%%%%%%%%%%%%%%%%%%%%%%%%%%
%%%%%%%%%%%%%%%%%%%%%%%%%%%%%%%%%%%%%%%%%%%%%%%%%%%%%%%%%%%%%%%%%%%%%%%%%%%%%

\smallskip
In summary, we have obtained an improved prediction for the light
$\cp$-even Higgs boson mass in the MSSM by combining the
FD result at the one- and two-loop level with an
all-order resummation of the leading and subleading logarithmic 
contributions from the top/stop sector obtained from solving the two-loop
RGEs. Particular care has been taken
to consistently match these two different types of corrections. The result,
providing the most precise prediction for $\Mh$ in the presence of large
masses of the scalar partners of the top quark, has been implemented
into the public code \fh. We have found a sizable effect of the
higher-order logarithmic contributions for 
$\MS \equiv \sqrt{\mste\mstz} \gsim 2 \tev$ which grows with increasing
$\MS$. 
In comparison with {\tt H3m}, which contains the
  \order{\alt\als^2} corrections to $\Mh$, we find that additional 
3-loop corrections of
\order{\alt^2\als,\alt^2} and also higher-loop corrections are both
important for a precise $\Mh$ prediction, amounting to effects of 
$\sim 7 \gev$ for $\MS = 10 \tev$ in our example.
%differences of $\sim 7 \gev$ are found between {\tt H3m} and \fh.
Finally,
we have shown that for sufficiently high $\MS$ the predicted
values of $\Mh$ reach about $126 \gev$ even for vanishing mixing in the
scalar top sector. As a consequence, even higher $\MS$ values are
disfavoured by the measured mass value of the Higgs signal.

%%%%%%%%%%%%%%%%%%%%%%%%%%%%%%%%%%%%%%%%%%%%%%%%%%%%%%%%%%%%%%%%%%%%%%%%%%%%%
%%%%%%%%%%%%%%%%%%%%%%%%%%%%%%%%%%%%%%%%%%%%%%%%%%%%%%%%%%%%%%%%%%%%%%%%%%%%%

%\subsection*{Acknowledgements}

\bigskip
{\bf Acknowledgements: } 
We thank 
H.~Haber, 
P.~Kant, 
P.~Slavich 
and 
C.~Wagner
for helpful discussions.
The work of S.H.\ was supported by the 
Spanish MICINN's Consolider-Ingenio 2010 Program under Grant MultiDark No.\ 
CSD2009-00064. 
The work of G.W.\ was supported by the Collaborative
Research Center SFB676 of the DFG, ``Particles, Strings, and the early
Universe".

%%%%%%%%%%%%%%%%%%%%%%%%%%%%%%%%%%%%%%%%%%%%%%%%%%%%%%%%%%%%%%%%%%%%%%%%%%%%%
%%%%%%%%%%%%%%%%%%%%%%%%%%%%%%%%%%%%%%%%%%%%%%%%%%%%%%%%%%%%%%%%%%%%%%%%%%%%%

\end{document}